\documentclass[11pt,a4paper]{article} 
\usepackage{geometry} 
\usepackage{authblk}

\usepackage{graphicx} 
\usepackage{epstopdf}
\usepackage{dcolumn}	 % Align table columns on decimal point
\usepackage{bm}		 
\usepackage{amssymb,amsmath}
\usepackage{latexsym}
\usepackage{color}
\usepackage[breaklinks,colorlinks,bookmarks=false,citecolor=blue,linkcolor=red,urlcolor=blue]{hyperref}

\graphicspath{{./}, {Figures/}}

\newcommand{\Eqref}[1]{Eq.~\eqref{#1}}
\newcommand{\Eqrefs}[2]{Eqs.~(\ref{#1},\ref{#2})}
\newcommand{\Figref}[1]{Fig.~\ref{#1}}

\newcommand{\bea}{\begin{eqnarray}}
\newcommand{\eea}{\end{eqnarray}}

\newcommand{\be}{\begin{equation}}
\newcommand{\ee}{\end{equation}}

\newcommand{\e}{{\mathrm{e}}}

\newcommand{\TK}{T_{\mathrm{K}}}

\newcommand{\HH}{{\cal{H}}}

\DeclareMathOperator{\diag}{diag}
\DeclareMathOperator{\median}{median}
\DeclareMathOperator{\width}{width}

\begin{document}

\title{Towards verified numerical renormalization group calculations}

% \author{Peter Schmitteckert}
% \institute{Lehrstuhl f{\"u}r Theoretische Physik I\\
% Physikalisches Institut\\
% Am Hubland\\
% Universit{\"a}t  W{\"u}rzburg\\
% 97074 W{\"u}rzburg\\
% Germany\\
% \email{Peter.Schmitteckert@physik.uni-wuerzburg.de}}

\author{Peter Schmitteckert}
\affil{Lehrstuhl f{\"u}r Theoretische Physik I, Physikalisches Institut, Am Hubland, Universit{\"a}t  W{\"u}rzburg, 97074 W{\"u}rzburg, Germany}

\date{\today}

\maketitle
\abstract{Numerical approaches are an important tool to study strongly correlated quantum systems.
However, their fragility with respect to rounding errors is not well studied and numerically verified
enclosures of the results are not available.
In this work we apply interval arithmetic to the well established numerical renormalization group scheme.
This extension enables us to provide a numerically verified NRG excitation spectrum.}

%\maketitle

\section{Introduction}

Wilson's numerical renormalization group scheme \cite{Wilson:RMP75,Wilson:Adv75,Hewson:93,CostiPruschke:RMP08}
is one of the most import numerical schemes in the field of strongly correlated quantum systems allowing
to track the full crossover from weak to strong coupling within the Kondo problem.

The Kondo model and the associated Kondo resonance can be seen as the prime examples
for correlated quantum systems. The Kondo problem itself can be traced back to the experiments
by de Haas  and van den Berg \cite{deHaas:P36} in the early 1930s which displayed an increase in resistivity
of noble metals like gold or silver. It took 30 years until Kondo \cite{Kondo:PTP64,Kondo:SSP69} could relate the increase of
resistivity to dynamical scattering at magnetic impurities. But it was only more than 30 years later
that Wilson could provide a rigorous solution of the Kondo model based on his
numerical renormalization group (NRG) technique.
A few years later Andrei \cite{Andrei:PRL80} and Vigman \cite{Wiegman:JETP80} could verify the numerical solution
of Wilson by a Bethe ansatz solution.
Besides its importance in describing magnetic impurities in metals it is also important in understanding
the transport properties of quantum dots \cite{GoldhaberGordon:PRL98} and often appears as effective model
in understanding correlated quantum systems, e.g. the dynamical mean field theory \cite{MetznerVollhardt:PRL89,Georges:RMP96}. 
For an overview see \cite{CostiPruschke:RMP08}.

Besides being used for decades  it was realized only recently \cite{PS:X2018}, that numerical rounding errors lead to the appearance
of a new fixed point within the NRG. Most strikingly this fixed point obeys a scaling law with respect to the precision 
of the underlying arithmetic and behaves like a typical physical fixed point. In this work we investigate, whether one can use interval
arithmetic to signal the breakdown of the numerics and to provide a scaling regime where the correctness of the numerical result can be guarantied.

%%%%%%%%%%%%%%%%%%%%%%%%%%%%%
\section{Interval arithmetic}
%%%%%%%%%%%%%%%%%%%%%%%%%%%%%
%%
Interval arithmetic was already introduced by Ramon Moore \cite{Moore:1966} in the 1960s
as an approach to bound rounding errors in mathematical computations. 
Within interval arithmetic \cite{Moore:1966,Alefeld:1983,boost,Filib_2006} one represents a number not by a single discretized floating point value.
Instead it is represented by two floating point values, a lower and an upper bound presenting an enclosure of
the the desired values $x$,
\begin{align}
  x \in \left[ \underline{x}, \overline{x} \right] \quad  \underline{x} \le \overline{x}  \label{eq:interval_def}.
\end{align}
In addition all operations $\star$ and functions $f(x)$ are extended on intervals in such a way
\begin{align}
   \left[ \underline{z}, \overline{z} \right] &=  \left[ \underline{x}, \overline{x} \right] \star  \left[ \underline{y}, \overline{y} \right] \;:\;
   \forall x \in  \left[ \underline{x}, \overline{x} \right] \wedge y \in  \left[ \underline{y}, \overline{y} \right] 
    \; x \star y \in \left[ \underline{z}, \overline{z} \right] \label{eq:interval_op}\\
    \left[ \underline{z}, \overline{z} \right] &= f\left( \left[ \underline{x}, \overline{x} \right]  \right) 
    \;:\: \forall x \in \left[ \underline{x}, \overline{x} \right]  \; f(x) \in \left[ \underline{z}, \overline{z} \right]. \label{eq:interval_f}
\end{align}
that the function values of $f(x)$ are contained in the result for all $x \in \left[ \underline{x}, \overline{x} \right]$,
and similarly for all operations $\star$. For example the addition of two intervals is now given by
\begin{align}
   \left[ \underline{x}, \overline{x} \right] +  \left[ \underline{y}, \overline{y} \right]
    &=      \left[ \underline{ \underline{x} + \underline{y}}, \overline{\overline{x} + \overline{y}} \right]
\end{align}
where $\underline{ \underline{x} + \underline{y}}$ is the sum $\underline{x} + \underline{y}$ rounded downwards on the level of the numerical precision,
while $\overline{\overline{x} + \overline{y}}$ is the sum $\overline{x} + \overline{y}$ rounded upwards.
Provided all operations are implemented with the necessary rounding modes interval arithmetic allows for an enclosure of the actual result.
However, in general it is not possible to obtain the smallest possible enclosure of \Eqrefs{eq:interval_op}{eq:interval_f}.
In general one should expect that interval arithmetic overestimates the actual numerical error in cases where one simply replaces
the floating point values by interval arithmetic in a given code. A prime example is given by the square function evaluated
on an interval containing zero, e.g. $[-1, 1]$
\begin{align}
 \left( [-1, 1]\right)^2   &=  [ 0, 1 ] \label{eq:square_f} \\
  [-1, 1] \cdot  [ -1, 1] &=  [ -1, 1 ] \label{eq:square_xx}  \,.
\end{align}
Since the square function maps a real value on a non-negative number we obtain \Eqref{eq:square_f} from the definition \Eqref{eq:interval_f}.
In contrast, according to \Eqref{eq:interval_op} the evaluation of the product \Eqref{eq:square_xx} has to include the negative part.
One should note, that the result of \Eqref{eq:square_xx} encloses the result of \Eqref{eq:square_f}. 
In the following we investigate the results of NRG simulations by simply replacing the floating point values
in \cite{PS:X2018} by an interval arithmetic.
%%
%%%%%%%%%%%%%%%%%%%%%%
\section{Kondo model}
%%%%%%%%%%%%%%%%%%%%%%
Here we follow precisely \cite{PS:X2018} in the description as well as the code used.
Note that the description in \cite{PS:X2018} is based on section {\tt VII} and {\tt VIII} of \cite{Wilson:RMP75}.

The Kondo model describes a local impurity coupled to a conduction band, where the
conduction band is transformed into spherical harmonics around the impurity and only the $s$-wave contributions are kept.
The remaining model of a spin impurity coupled to a half infinite chain is then discretized on a logarithmic scale.
The system is then tridiagonalized leading to the following form:
\begin{align}
	\HH_M &= J \vec{\hat{S}} \vec{\hat{s}}_1 \,+\, \sum_{n=2}^{M}  t_{n-1}  \sum_\sigma \hat{c}^\dagger_{n,\sigma}  \hat{c}^{}_{n-1,\sigma} \,+\, \text{h.c.} \,. \label{eq:KondoModel}
\end{align}

Here we follow the usual convention of $\vec{\hat{S}}$ being the $SU(2)$ spin operator of the impurity,
 $\vec{\hat{s}}_1$ is the spin operator of the first conduction band site, $\hat{c}^{}_{n,\sigma}$ ($\hat{c}^\dagger_{n,\sigma}$)
is the annihilation (creation) operator for a conduction band fermion with spin $\sigma$ in energy shell  $n$. 
$J$ denotes the Heisenberg exchange coupling and 
\begin{align}
	t_n = t \Lambda^{(n-1)/2} \label{eq:tn}
\end{align}
is taken in its most simplistic form ignoring any corrections stemming from the original band structure
as we are interested in the low energy physics only. For a justification for this Hamiltonian we refer to 
excellent articles by Wilson \cite{Wilson:RMP75,Wilson:Adv75}.  
However, for this work it is sufficient to know that $\HH$ \eqref{eq:KondoModel} describes a single spin
coupled to a 1D like tight binding chain where the $n$-th site represents the physics at
energy scale $t\, \Lambda^{(n-1)/2}$.

The most prominent property of the Kondo model is the flow from a weak coupling regime represented
by a spin impurity coupled weakly to the conduction band consisting of $M$ sites, to a strong coupling regime, where 
the spin impurity forms a singlet with the conduction band fermions leading to a singlet weakly coupled
to an effective conduction band consisting of $M-1$ sites. For a discussion of the parity effect with respect to $M$
we refer to \cite{PS:X2018}.

In order to obtain this flow the NRG starts with a system consisting of the impurity spin and the first
conduction band site, that is with Hamiltonian \eqref{eq:KondoModel} setting $M=1$.
One then iteratively increases the number of conduction band sites by one. Since the associated Hilbert space
will increase by a factor of four in each step one has to introduce a truncation scheme. Within the NRG scheme
one truncates the Hilbert space after each diagonalization to the $m$ eigen states lowest in energy.
In addition one performs a shift of the eigen values such that the lowest eigen value is zero, $E_0=0$.

\begin{figure}[t]
 \centerline{\includegraphics[width=0.8\textwidth]{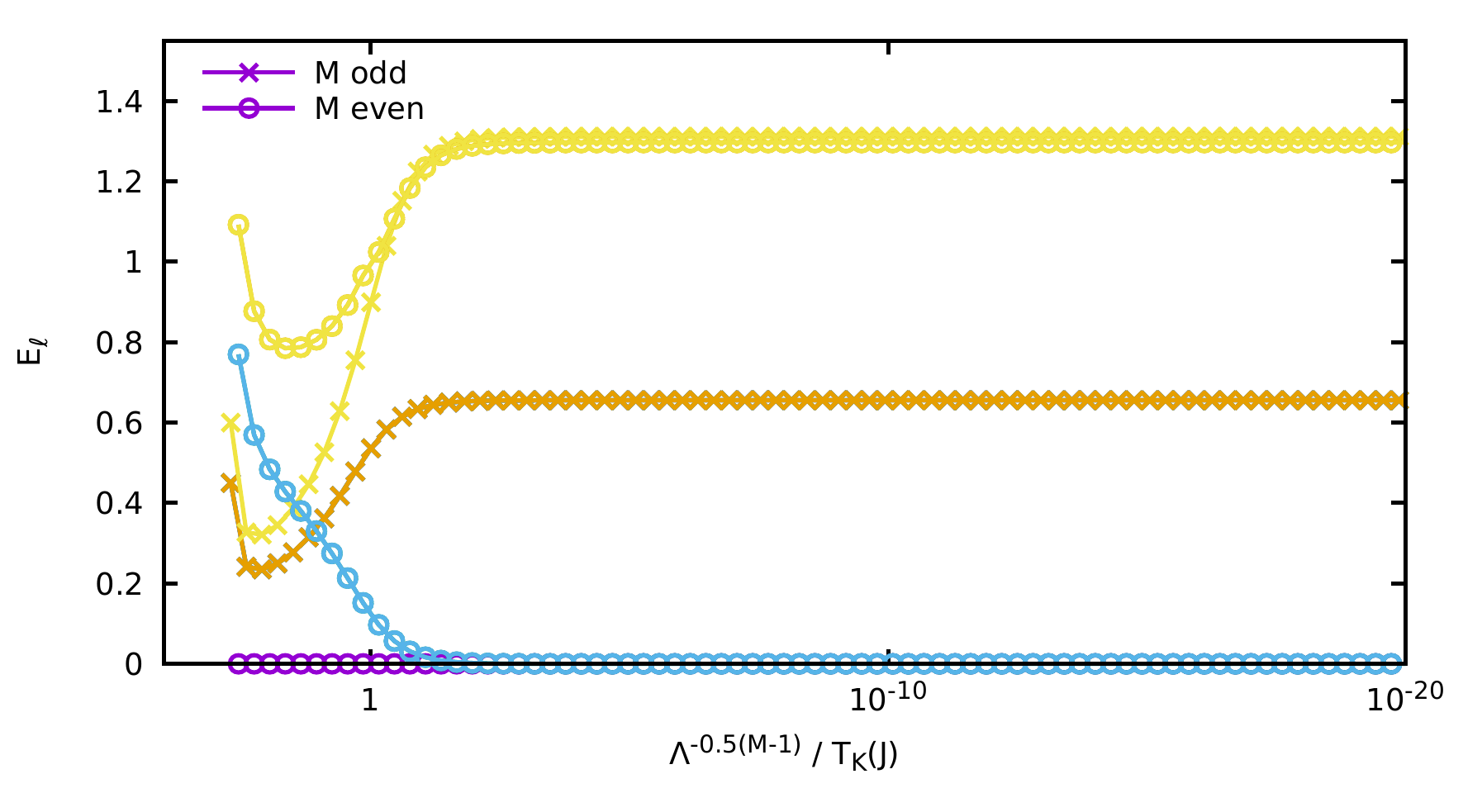}}
 \caption{\label{fig:RGFlowKondo} The RG flow of the low energy spectrum for a Kondo model with $J=0.6$,
   $\Lambda=2.0$, and $m=2000$, where the $N$ and $S^z$ conservation is explicitly enforced. 
   The results are split into odd conduction band sites ($\circ$) and even conduction band sites ($\times$).
   }
\end{figure}

Since the low energy scale of the  Hamiltonian \eqref{eq:KondoModel}  decreases by $\Lambda^{-1/2}$ in each iteration
step one rescales the Hamiltonian in order to keep the excitation energies of order one
\begin{align}
	\widetilde{\HH}_M                 &= \Lambda^{M/2} \HH_M \label{eq:KondoTilde}  \,.
\end{align}

In this form one can now investigate the flow of the spectrum. Of course, in order to get the corresponding physical scale one has
to undo the scaling. 
The striking feature of the Kondo model is the appearances of a scale $\TK$
\begin{equation}
	\TK = D \sqrt{J/D} \,\e^{-D/J} \label{eq:TK} \,
\end{equation}
with the $D=4t$ the band width of the Hamiltonian \Eqref{eq:KondoModel} and $t$ the band hopping element \Eqref{eq:tn}.

As an example we provide in \Figref{fig:RGFlowKondo} the  RG flow for a Kondo system with a spin coupling of $J=0.6$.
There we show the flow of the five lowest excitation energies vs.\ the energy scale 
$t_M = \Lambda^{(M-1)/2}$ in units of $\TK$. In this computation the particle number and the $S^z$ component of the 
total spin were explicitly conserved by working with a block matrix representation of the Hamiltonian
and we truncated the Hilbert space to at most $m=2000$ states.
One clearly observes the crossover regime at $\TK$ and the flow towards the strong coupling regime at low energies.
For a detailed description we refer to \cite{Wilson:RMP75,Wilson:Adv75,PS:X2018}.

\begin{figure}[t]
 \centerline{\includegraphics[width=0.8\textwidth]{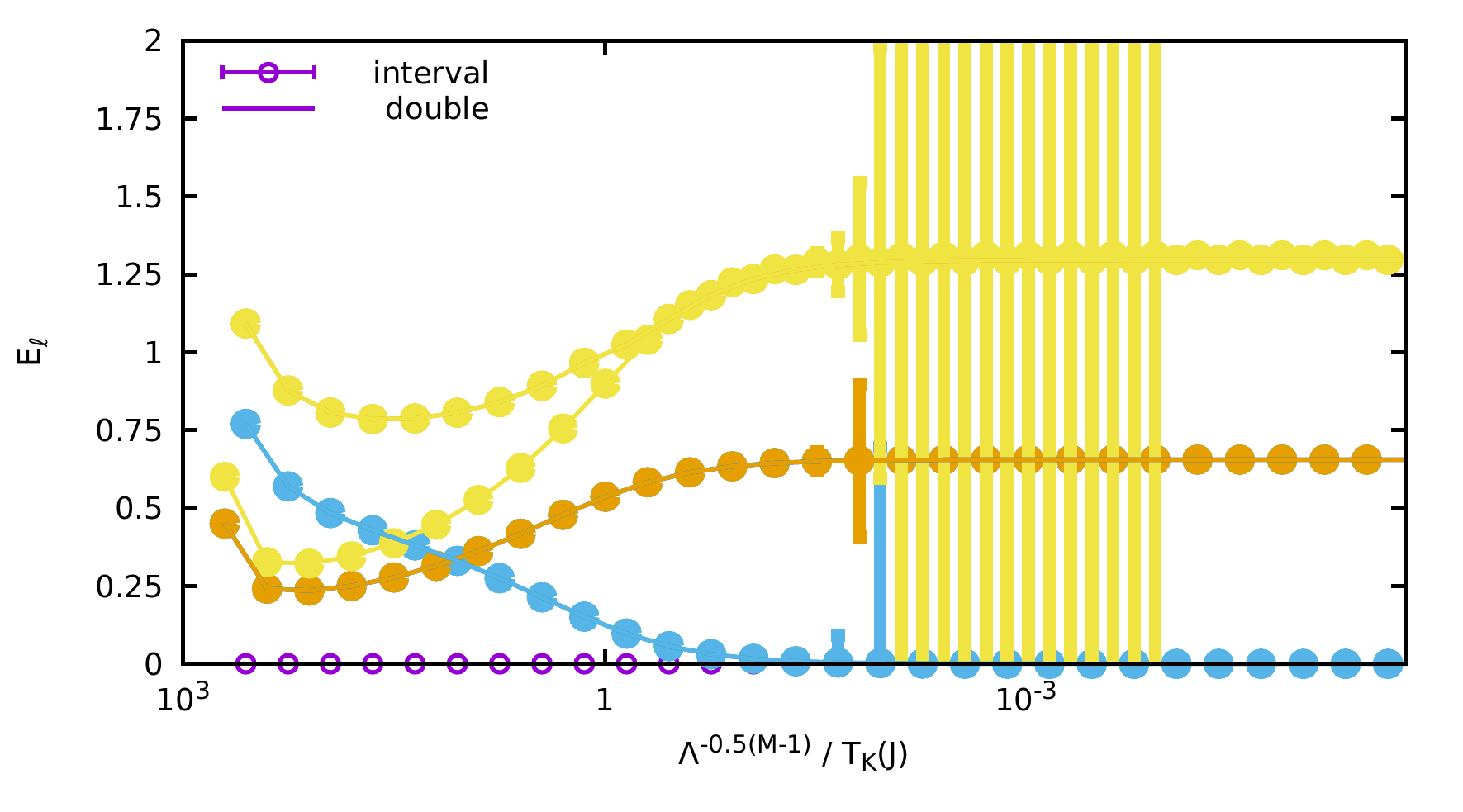}}
 \caption{\label{fig:RGFlowKondoIL2} The RG flow of the low energy spectrum for a Kondo model with $J=0.6$
   $\Lambda=2.0$ and $m=2000$, where the $N$ and $S^z$ conservation is explicitly enforced. 
   The lines correspond to the results to a standard NRG using {\tt double} arithmetics as presented in \Figref{fig:RGFlowKondo}.
   The $\circ$ correspond to the median of the obtained excitation energies. The error bars correspond to the width
   of the excitation energy intervals. In the left part of the figures, the error is smaller then the symbol size, while in the right part of
   the figure the errors are much larger than any physical scale.
   }
\end{figure}

\section{NRG with interval arithmetics}
In order to track rounding issues we performed the following change to our code.
We replaced the data type from {\tt double} to {\tt interval<double>} which is straightforward
everywhere with the exception of the diagonalization of the Hamiltonian matrix.
Here we took the simple approach of extracting a median matrix
\begin{align}
 \median\left( \left[ \underline{x}, \overline{x} \right] \right) &= \left( \underline{x} + \overline{x} \right)/2 \\
 h_{x,y} &= \median( H_{x,y} ) \,.
\end{align}
We then diagonalize matrix $h$ in a standard way, as its elements are of type {\tt double}.
We then take the resulting transformation matrix $U$ as the transformation matrix
and obtain the new energy eigenvalues as the diagonal elements obtained by a base transformation of $H$ via $U$:
\begin{align}
 E &= \diag\left( U^+ \cdot H \cdot U \right) \,.
\end{align}
We would like to point out that these changes are rather simple and should be applicable to any NRG implementation.

In result we now obtain an interval for each energy value where the width of the interval
\begin{align}
 \width\left( \left[ \underline{x}, \overline{x} \right] \right) &= \overline{x} - \underline{x} 
\end{align}
provides an error bar for the calculation.

\begin{figure}[t]
 \centerline{\includegraphics[width=0.8\textwidth]{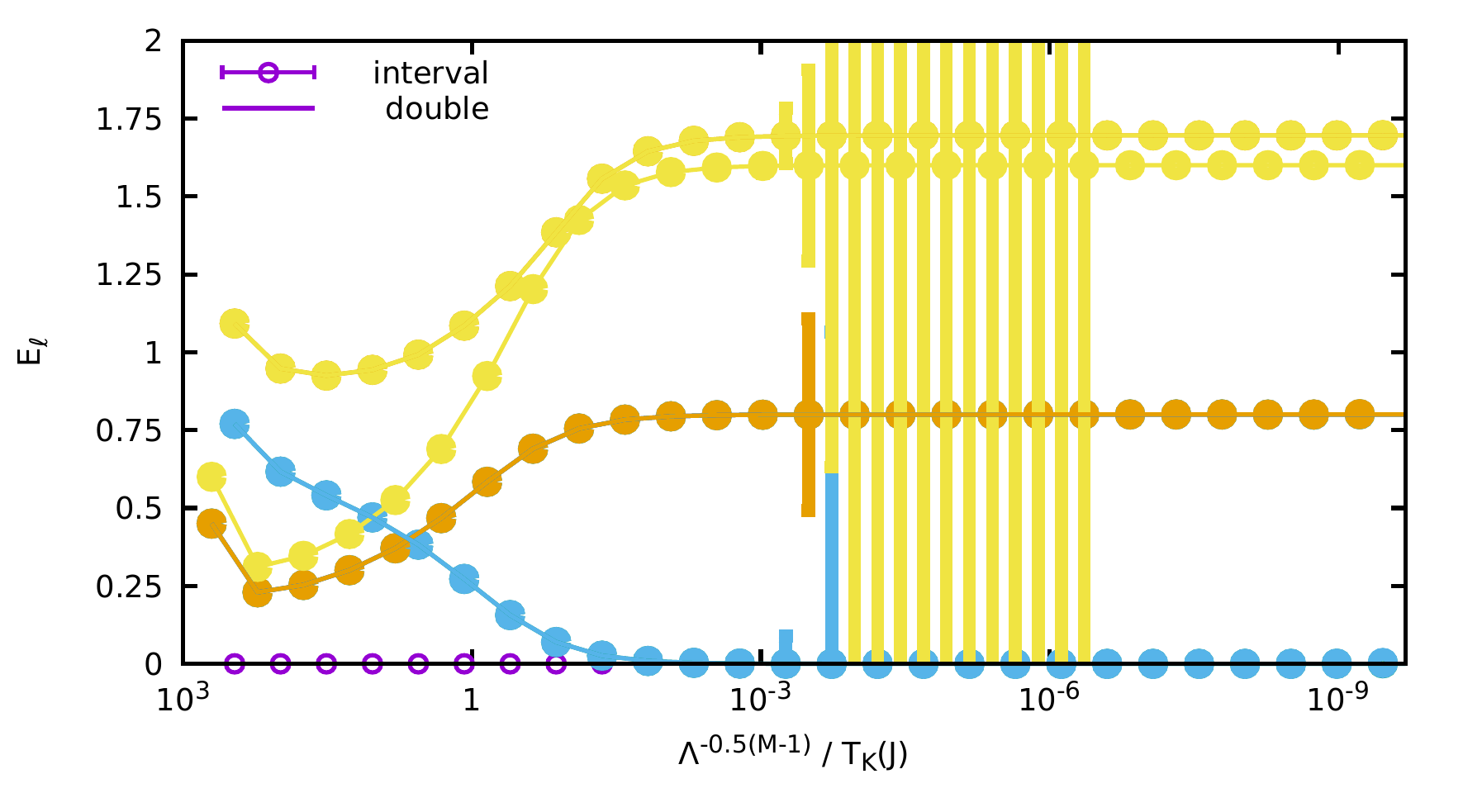}}
 \caption{\label{fig:RGFlowKondoIL3} The RG flow of the low energy spectrum for a Kondo with $J=0.6$
   $\Lambda=3.0$ and $m=500$, where the $N$ and $S^z$ conservation is explicitly enforced. 
   The lines correspond to the results to a standard NRG using {\tt double} arithmetics.
   The $\circ$ correspond to the median of the obtained excitation energies. The error bars correspond to the width
   of the excitation energy intervals.
   }
\end{figure}

In \Figref{fig:RGFlowKondoIL2} we provide the results for an interval version of the results presented in \Figref{fig:RGFlowKondo}.
\Figref{fig:RGFlowKondoIL3} corresponds to the same system with $\Lambda=3.0$ and $m=500$.
The first observation from these results is that one can actually perform an interval version of the NRG scheme and the median of the
energy eigenvalue intervals corresponds to those of an NRG with standard $\tt double$ arithmetics during the complete crossover
to the strong coupling regime and verified that this crossover is not due to rounding errors.

\begin{figure}[t]
 \centerline{\includegraphics[width=0.8\textwidth]{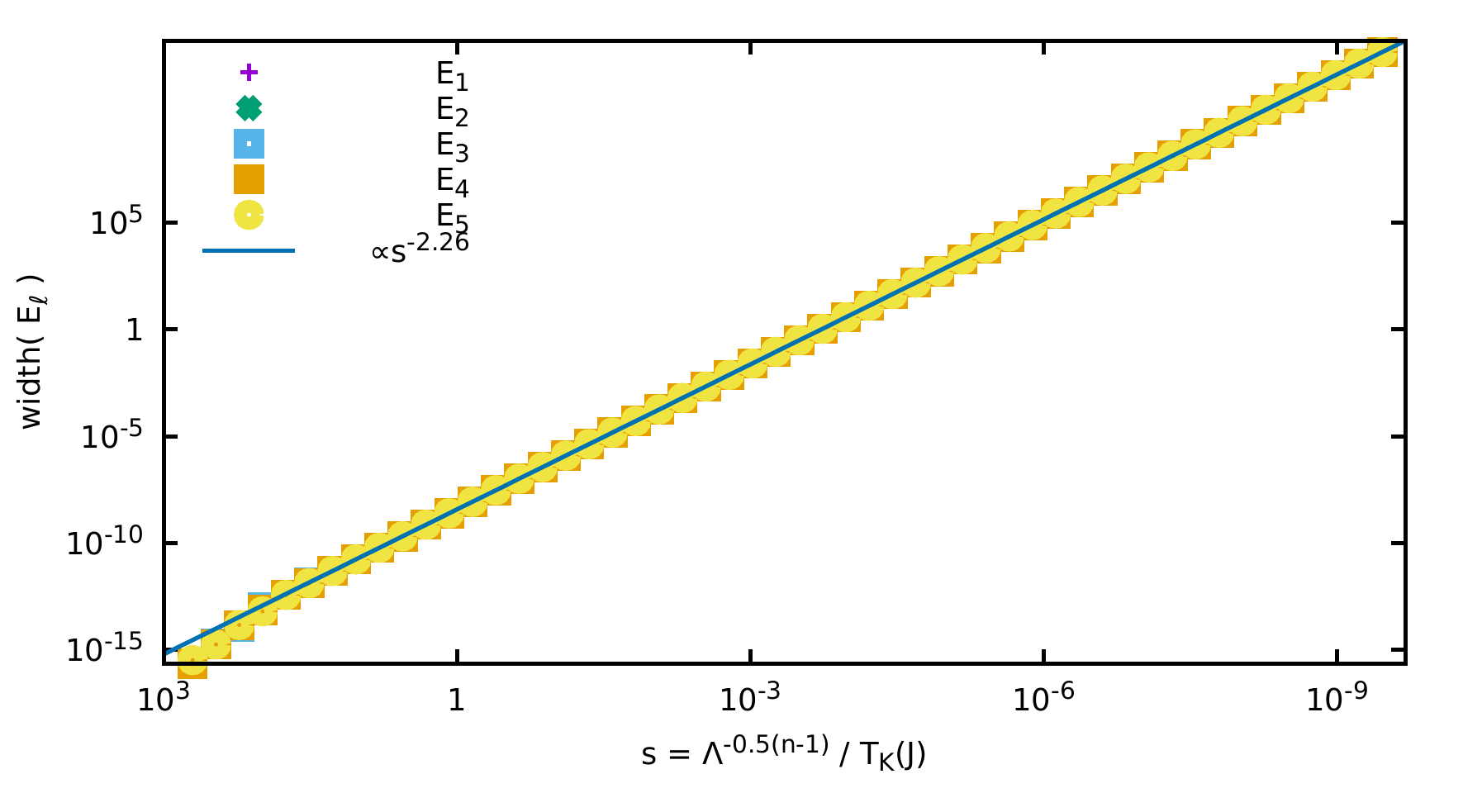}}
 \caption{\label{fig:RGFlowKondoIL3err} The width of the energy eigenvalue intervals RG corresponding to the results of \Figref{fig:RGFlowKondoIL3}.
    The line is a power law fit resulting in an exponent of $\alpha \approx -2.26$.}
\end{figure}

In \Figref{fig:RGFlowKondoIL3err} we provide the width of the energy eigenvalue intervals corresponding to the results in \Figref{fig:RGFlowKondoIL3}.
Here we witness a power law increase of the width of the energy eigenvalue intervals with respect to the low energy scale. 
Once the width of the energy intervals gets larger than the actual level splitting the interval arithmetic signals the end of a numerically verified
spectrum of the NRG scheme. We would like to point out that this does not imply that a corresponding NRG calculation within {\tt double}
arithmetic breaks down. As pointed out in the introduction the interval scheme employed here will overestimate the corresponding numerical error.

%%%%%%%%%%%%%%%%%%%%%%%%%%%%%%%%%%%%%%%%%%%%%%%%%%%%%%%%%%%%%%%%%%%%%%%%%%%%%%%%
\section{Summary \& Outlook}
%%%%%%%%%%%%%%%%%%%%%%%%%%%%%%%%%%%%%%%%%%%%%%%%%%%%%%%%%%%%%%%%%%%%%%%%%%%%%%%%

In this work we extended Wilson's NRG scheme to interval arithmetics which allows us to 
provide a numerical guarantee on the obtained spectrum, provided the width of the energy
eigenvalue intervals do not signal a break down. We presented results keeping $m=2000$ states after
each NRG step. This is sufficiently large enough to demonstrate, that one is not restricted to toy calculations.
The approach presented provides a measure to assure that the result is not dominated by the finite 
precision arithmetic as the parity breaking fixed point in \cite{PS:X2018}.
Within the scheme employed here we observe a power law increase of the width of the energy eigenvalue intervals with respect to the inverse RG scale
which currently limits the applicability of the verified NRG scheme to not too small energy scales.
We therefore had to choose a rather large Kondo coupling, $J=0.6$. For significantly smaller couplings the error would blow up before
the strong coupling regime is reached.
It remains an open question whether this could be significantly improved by the application of an interval version of the diagonalization
tailored to obtain smaller bounds on the eigen values and therefore leading to a verified NRG scheme applicable in the whole parameter regime.
Alternatively one could apply a multi-precision interval library in a similar way as in \cite{PS:X2018} to extend the range
of applicability of the approach presented in this work.

\section*{Acknowledgement}
%\begin{acknowledgement}
%
 The numerics is performed using the Eigen 3 library \cite{eigen3}
 and {\tt g++} from the Gnu compiler collection \cite{gcc}.
 We used the interval library from {\tt boost.org} \cite{boost} and {\tt FILIP++} \cite{Filib_2006}.
 Both libraries provide very similar results.
%
%\end{acknowledgement}

%%%%%%%%%%%%%%%%%%%%%%%%%%%%%%%%%%%%%%%%%%%%%%%%%%%%%%%%%%%%%%%%%%%%%%%%%%%%%%%%
%%%  BIBLIOGRAPHY generated with Jabref, .bbl should be pasted here 
%%%%%%%%%%%%%%%%%%%%%%%%%%%%%%%%%%%%%%%%%%%%%%%%%%%%%%%%%%%%%%%%%%%%%%%%%%%%%%%%

%\bibliographystyle{abbrvdin}
%\bibliography{References}

%%%%%%%%%%%%%%%%%%%%%%%%%%%%%%%%%%%%%%%%%%%%%%%%%%%%%%%%%%%%%%%%%%%%%%%%%%%%%%%%
%%%%%%%%%%%%%%%%%%%%%%%%%%%%%%%%%%%%%%%%%%%%%%%%%%%%%%%%%%%%%%%%%%%%%%%%%%%%%%%%
%%%%%%%%%%%%%%%%%%%%%%%%%%%%%%%%%%%%%%%%%%%%%%%%%%%%%%%%%%%%%%%%%%%%%%%%%%%%%%%%
\end{document}